\documentclass[letter]{aa}

\usepackage{amsmath,amsfonts,amssymb} 
\usepackage{bm} 
\usepackage[table,  dvipsnames]{xcolor} 
\usepackage{hyperref} 
\usepackage{nicefrac} 
\usepackage{placeins}
\usepackage{microtype}

\usepackage[varg]{txfonts} 
\usepackage[T1]{fontenc}

\hypersetup{
	pdftitle = {Substructure in externally irradiated protoplanetary disks --- Part I: 2D radiative modeling},
	pdfauthor = {A. Ziampras, L. Qiao, T. J. Haworth},
	pdfsubject = {},
	colorlinks = {true},
	linkcolor = [rgb]{0,0.35,0.7},
	citecolor = [rgb]{0,0.35,0.7},
	filecolor = [rgb]{0.61,0,0},
	urlcolor = [rgb]{0.61,0,0},
}

\usepackage{graphicx}  
\graphicspath{{./}}

\newcommand{\D}[2]{\frac{\text{d}{#1}}{\text{d}{#2}}}

\newcommand{\G}{\text{G}}
\newcommand{\Mstar}{M_\star}
\newcommand{\Lstar}{L_\star}

\newcommand{\Msun}{\mathrm{M}_\odot}
\newcommand{\Lsun}{\mathrm{L}_\odot}

\newcommand{\Rgas}{\mathcal{R}}
\newcommand{\cs}{c_\mathrm{s}}

\newcommand{\OmegaK}{\Omega_\mathrm{K}}

\newcommand{\kappaR}{\kappa_\mathrm{R}}
\newcommand{\kappaP}{\kappa_\mathrm{P}}
\newcommand{\cv}{c_\mathrm{v}}
\newcommand{\rhomid}{\rho_\mathrm{mid}}
\newcommand{\sigmaSB}{\sigma_\mathrm{SB}}
\newcommand{\vel}{\bm{u}}

\newcommand{\lrad}{l_\mathrm{rad}}
\newcommand{\Sigmag}{\Sigma_\mathrm{g}}
\newcommand{\Sigmad}{\Sigma_\mathrm{d}}
\newcommand{\velg}{\vel_\mathrm{g}}
\newcommand{\veld}{\vel_\mathrm{d}}

\newcommand{\Lext}{L_\mathrm{ext}}
\newcommand{\dext}{d_\mathrm{ext}}
\newcommand{\Fext}{F_\mathrm{ext}}
\newcommand{\vecFext}{\bm{F}_\mathrm{ext}}

\newcommand{\pluto}{\texttt{PLUTO}}

\newcommand{\optool}{\texttt{optool}}

\begin{document}

\title{Substructure in externally irradiated protoplanetary disks}
\subtitle{I. spirals and rings in two-dimensional radiation hydrodynamics}
\titlerunning{Spirals and rings in externally irradiated disks}

\author{
	Alexandros~Ziampras\thanks{E-mail: a.ziampras@lmu.de}\inst{\ref{inst1},\ref{inst2},\ref{inst3}}
	\and Lin~Qiao\inst{\ref{inst2}}
	\and Thomas~J.~Haworth\inst{\ref{inst2}}
}

\institute{
	Ludwig-Maximilians-Universit{\"a}t M{\"u}nchen, Universit{\"a}ts-Sternwarte, Scheinerstr.~1, 81679 M{\"u}nchen, Germany \label{inst1}
	\and Astronomy Unit, School of Physics and Astronomy, Queen Mary University of London, London E1 4NS, UK \label{inst2}
	\and Max Planck Institute for Astronomy, Königstuhl 17, 69117 Heidelberg, Germany \label{inst3}
}

\date{\today}

\abstract{
	It is known that the external irradiation of protoplanetary disks by nearby massive stars can result in mass loss that impacts the disk evolution, however the dynamical impact of external irradiation upon the disk itself has not been explored in detail.
	We aim to investigate the dynamical effect of asymmetric external irradiation on the structure of such disks.
	We perform two-dimensional multi-fluid radiation hydrodynamical simulations of protoplanetary disks subject to external irradiation using the \pluto{} code, with external irradiation modeled as a plane-parallel flux and a simplified nonaxisymmetric heating rate corresponding to the thermal reemission from hot material within the region marginally optically thick to the external irradiation.
	We find that a nearby massive star can, under certain conditions, induce significant dynamical effects on a protoplanetary disk, including a shadowed region, pronounced spiral arms in gas, and rings and gaps in dust. The dynamics are caused by the temperature asymmetry driven and maintained by external irradiation, akin to the well-established mechanism of shadow-induced spirals and rings in disk with shadowing from their inner regions.
	Our results show that if an external temperature asymmetry can be induced it can have a significant dynamical impact on the disk itself (in addition to the well-studied mass loss and truncation effects due to external irradiation), possibly even driving substructure. This prompts further investigation with detailed, dynamical radiative transfer models.
}


\bibpunct{(}{)}{;}{a}{}{,}

\maketitle

\section{Introduction}
\label{sec:introduction}
The dynamical evolution of protoplanetary disks is key to understanding planet formation. Disks exhibit many interesting dynamical processes, including decoupled dust/gas dynamics \citep[e.g.,][]{Weidenschilling-1977}, planet--disk interactions \citep{kley-nelson-2012} gravito- and pure hydrodynamic instabilities \citep{2016ARA&A..54..271K}, magneto- and radiation driven flows and instabilities \citep[e.g.,][]{1991ApJ...376..214B, 2013ApJ...769...76B}. Huge strides have been made towards understanding these both from a theoretical standpoint and observationally thanks to high resolution, high sensitivity facilities such as VLT/SPHERE and ALMA, which have revealed substructures such as gaps, rings, spirals and discrete clumps \citep[e.g.,][]{alma-etal-2015a, andrews-etal-2018,huang-etal-2024} that are predicted to arise from dynamical processes \citep[e.g.,][]{2015MNRAS.453L..73D}.

There is growing interest in the environmental impacts upon planet-forming disks in star forming regions. For example, infall of material may replenish the disk \cite[e.g.,][]{2024A&A...683A.133G, huhn-dullemond-2025} and stellar fly-bys can truncate disks \citep[e.g.,][]{clarke-pringle-1993,heller-1995,cuello-etal-2023}. There is also the process of ``external photoevaporation'', where UV radiation (particularly from massive stars) heats and drives winds from the outer regions of disks \citep[e.g.,][]{yorke-welz-1996,WinterHaworth2022}. Most research into external photoevaporation has been focused on the winds and mass loss rates \citep[e.g.,][]{Johnstone-etal-1998, Adams-etal-2004, haworth-etal-2018, haworth-etal-2023, Aru-etal-2024}, the implications that has for disk evolution and lifetimes \citep[e.g.,][]{Scally-Clarke-2001, Winter-etal-2018, ConchaRamirez-etal-2019, Coleman-Haworth-2022}, implications for planet formation \citep[e.g.,][]{Winter-etal-2022, Qiao-etal-2023, Hallatt-Lee-2025, qiao-etal-2026} and trying to understand the prevalence and timescales of given levels of irradiation \citep[e.g.,][]{Fatuzzo-Adams-2008, Qiao-etal-2022, Anania-etal-2025}. However, there are other effects of external irradiation that have been considered, including upon the disk chemistry \citep{2013ApJ...766L..23W, Boyden-Eisner-2023, 2025MNRAS.537..598K}.

An aspect of external disk irradiation that has not been considered in detail is the dynamical impact of external heating upon the disk itself. Shadowing of the outer disk by the inner disk has been shown to result in temperature asymmetries that lead to dynamical evolution \citep{montesinos-etal-2016,cuello-etal-2019,su-bai-2024,qian-wu-2024,zhang-zhu-2024,ziampras-etal-2025b,zhu-etal-2025}. When a disk is externally irradiated and subject to asymmetric heating there may similarly be stimulated dynamical consequences, that have an impact in addition to the mass loss in the wind, though this is yet to be studied. 

In this paper we use vertically integrated two-dimensional (i.e., coplanar with the disk) multi-fluid radiation hydrodynamical simulations with \pluto{} to make the first exploration of the dynamical impact of asymmetric externally irradiation upon the disk. Our physical framework and setup largely follows that of \citet{ziampras-etal-2025b}, with the addition of a nonaxisymmetric external irradiation source incident edge-on on the disk, with all additions and modifications detailed in Appendix~\ref{sec:physics-numerics}. In Sect.~\ref{sec:results} we present our results, followed by a discussion in Sect.~\ref{sec:discussion} and a summary in Sect.~\ref{sec:summary}.

\section{Results}
\label{sec:results}

\begin{figure*}[t]
	\centering
	\includegraphics[width=.95\textwidth]{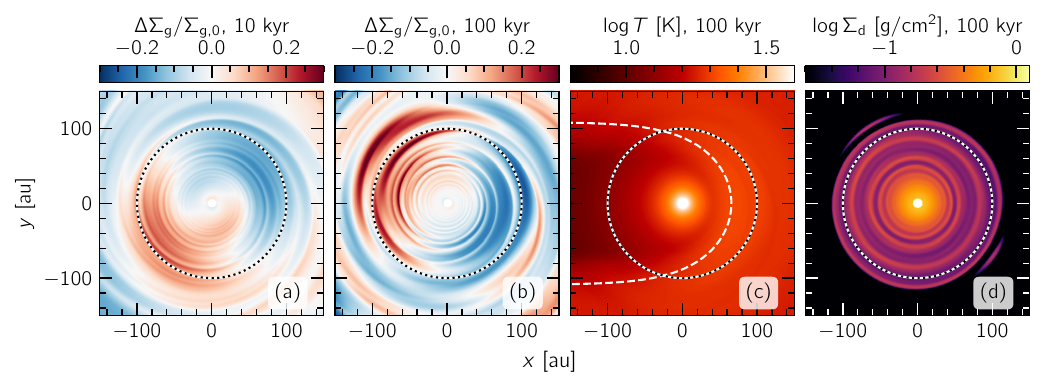}
    \vspace{-0.4cm}
	\caption{Heatmaps of the perturbed gas density $\Delta\Sigmag/\Sigma_\text{g,0}$ (panel \emph{b}), temperature $T$ (\emph{c}) and mm-grain dust density $\Sigmad$ (\emph{d}) after 100\,kyr of evolution in our model with external irradiation, showing prominent spiral structures in gas, the disk's colder, shadowed farside, and rings in the dust distribution.. Gas density perturbations are also shown at an earlier state (10\,kyr) on panel \emph{a}. A black--white dotted circle marks the exponential tapering radius of the disk at 100\,au. The dashed line in the middle panel indicates the $\tau=1$ surface towards the external source (here at a distance of 0.1\,pc towards the right, with a luminosity of $2\times10^5\,\Lsun$), albeit computed with several underlying simplifications in Appendix~\ref{sec:physics}.}
	\label{fig:fiducial-heatmaps}
\end{figure*}
\begin{figure}
	\centering
	\includegraphics[width=.9\columnwidth]{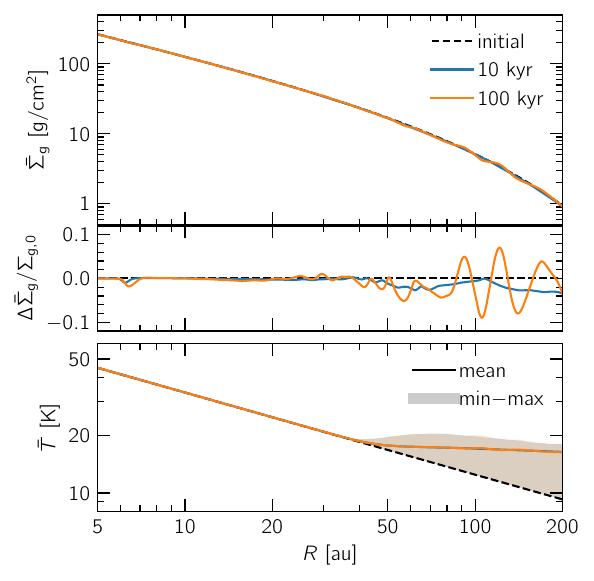}
    \vspace{-0.2cm}
	\caption{Azimuthally averaged radial profiles of the gas surface density (top) and temperature (bottom) as a function of time. Top: the gas density perturbations, while at the few \% level, grow throughout the simulation runtime due to the spirals induced by external irradiation. Bottom: the shaded region marks the temperature variation between the irradiated (upper bound) and shadowed (lower bound) sides of the disk.}
	\label{fig:radial-profiles}
\end{figure}
\begin{figure}
	\centering
	\includegraphics[width=\columnwidth]{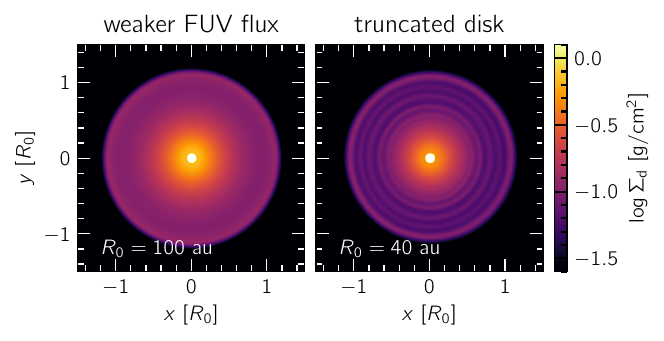}
	\caption{Dust surface density heatmaps after 100 orbits at $R_0$ in two additional models. Left: placing the external source at a distance of 1\,pc rather than the fiducial 0.1\,pc led to no noticeable substructure after 100 orbits (100\,kyr). Right: a less massive, smaller disk shows substructure in the form of circular rings after 100 orbits at $R_0=40$\,au (25\,kyr), albeit less prominent than in the fiducial model.}
	\label{fig:appendix-heatmaps}
\end{figure}
\begin{figure}
	\centering
	\includegraphics[width=\columnwidth]{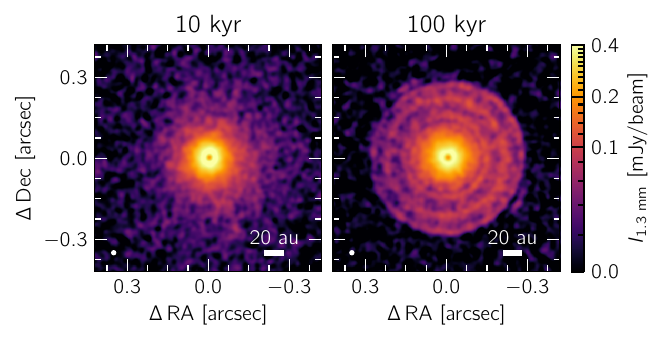}
	\caption{Synthetic ALMA observations at 1.3\,mm of our model, loosely mimicking a source at 400\,pc distance, after 10 (left) and 100\,kyr (right) of evolution. Rings only become visible after $\sim\!30$\,kyr for this model.}
	\label{fig:synthobs}
\end{figure}
The disk state for our fiducial, strongly irradiated model after 100\,kyr of evolution is shown in Fig.~\ref{fig:fiducial-heatmaps}, where we plot the perturbed gas surface density, the temperature, and the mm-grain dust surface density, from left to right. Azimuthally averaged profiles of the gas surface density, its radial perturbations, and the disk temperature are shown in Fig.~\ref{fig:radial-profiles}. From these two figures, it becomes clear that the strong nonaxisymmetric heating from the external source has a profound and immediate effect on the disk structure in multiple ways.

The temperature structure (panel \emph{c} of Fig.~\ref{fig:fiducial-heatmaps}) shows a clear asymmetry between the irradiated (right) and shadowed (left) sides of the disk, with the shadowed region roughly bounded by the $\tau=1$ surface towards the external source (dashed line in panel \emph{c} of Fig.~\ref{fig:fiducial-heatmaps}, see Eq.~\eqref{eq:tau_ext}). The boundary between the internally and externally heated regions at around 40\,au is also visible by eye in the figure. This temperature structure is established very early on in the simulation (within the first few orbits at 100\,au) and persists throughout, with only minor fluctuations due to the dynamics induced by this very heating asymmetry. From the bottom panel of Fig.~\ref{fig:radial-profiles}, we can see that the temperature contrast between the irradiated and shadowed sides of the disk is such that the temperature on the shadowed side remains close to the initial profile (i.e., heated by irradiation from the central star), whereas the irradiated side features a roughly constant temperature of $\sim\!17$\,K beyond $\sim\!40$\,au. This flattening of the temperature profile is in agreement with previous studies of externally irradiated disks \citep[e.g.,][]{haworth-2021}.

As a result of the nonaxisymmetric temperature, the gas surface density $\Sigmag$ (panel \emph{a} and \emph{b} of Fig.~\ref{fig:fiducial-heatmaps}) features prominent spiral arms of amplitude $\Delta\Sigmag/\Sigma_\text{g,0}\gtrsim 0.2$ developing within the first 10\,kyr, that persist throughout the simulation runtime. These spirals are launched due to the azimuthal modulation of the radial pressure gradient caused by the external heating source, with the mechanism detailed in \citet{ziampras-etal-2025b} and \citet{zhu-etal-2025}.
As detailed in these studies \citep[see also][]{su-bai-2024,qian-wu-2024}, the spiral arms carry an angular momentum flux that is deposited at the edge of the shadowed region with a radial spacing set by the sound speed profile of the disk \citep{ziampras-etal-2025b}. This process results in weak but persistent radial modulations in the gas surface density (see also Fig.~\ref{fig:radial-profiles}), and ultimately leads to the formation of rings and gaps in the dust surface density (right panel of Fig.~\ref{fig:fiducial-heatmaps}) due to dust trapping within the gas pressure maxima. This finding is in line with the aforementioned studies of shadow-induced spirals and rings, and suggests that external irradiation can also lead to similar substructures in protoplanetary disks.

At the same time, the disk seems to develop a lopsided structure, with the shadowed side away from the external source being denser than the irradiated side. This is likely related to the radial temperature profile being flatter on the irradiated side but maintaining its powerlaw shape within the shadow (see also Fig.~\ref{fig:radial-profiles}), which leads to a weaker radial pressure gradient on the irradiated side. This in turn results in a faster azimuthal flow there due to the gas being less pressure-supported and thus closer to its Keplerian speed, and culminates in a redistribution of mass towards the shadowed side of the disk. This behavior is also evident in the rings in dust surface density, which show a slight eccentricity with their apastron pointing towards the shadowed side of the disk.

We stress that the degree of dynamical evolution, as discussed at length in Sect.~\ref{sec:physics}, is quite sensitive to the intensity and geometry of the heating rate attributed to the external source. To address this, we carried out two additional models, which highlight that our fiducial model, while dynamically the most interesting, should be interpreted as an upper limit to the effects of external irradiation unless demonstrated otherwise by more self-consistent 3D radiative transfer models. The additional configurations correspond to a) a weaker external irradiation flux by a source at $\dext=1$\,pc (effectively reducing $\Fext$ by a factor of 100 to $10^3\,G_0$), and b) a smaller, less massive disk with a total gas mass of $\approx\!0.01\,\Msun$ that is truncated at 40\,au. The latter is motivated by the expectation that external photoevaporation, being a continuous process, will likely truncate the disk and reduce its mass over time, and thus this model can be thought of as a later evolutionary stage of the fiducial model.

The resulting disk states after 100 orbits at $R_0$ (100\,kyr for the left panel, 25\,kyr for the right panel) are shown in Fig.~\ref{fig:appendix-heatmaps}, where we plot the dust surface density similar to Fig.~\ref{fig:fiducial-heatmaps}. We find that no substructure whatsoever is visible in the dust distribution of the model with $\dext=1$\,pc, while the truncated disk features several rings similar to the fiducial model, albeit more circular and with a weaker contrast.

Finally, we produced synthetic ALMA observations of our fiducial model after 10 and 100\,kyr of evolution by computing the intensity at $\lambda=1.3$\,mm ($\nu=230$\,GHz) as
\begin{equation}
	\label{eq:I_nu}
	I_\nu = B_\nu(T)\left(1 - e^{-\kappa_\nu\Sigmad}\right), \quad \kappa_\text{1.3\,mm} = 3.06\,\text{cm}^2/\text{g},
\end{equation}
with $\kappa_\text{1.3\,mm}$ computed using \optool{} \citep{dominik-etal-2021} with DIANA-standard composition. The source was placed at a distance of 400\,pc. We then convolved the resulting intensity maps with a Gaussian beam of FWHM 20\,mas and overlaid Gaussian noise with an RMS of 10\,$\mu$Jy/beam. The resulting images are shown in the left panel of Fig.~\ref{fig:synthobs}, with the rings clearly visible in continuum emission on the right. For our fiducial model, while the gas density and temperature have already developed features within the first 10\,kyr of evolution (see Fig.~\ref{fig:fiducial-heatmaps}), rings in the mm dust continuum required at least $\sim\!30$\,kyr to become visible.

\section{Discussion}
\label{sec:discussion}

We briefly highlight the relevance of our findings in the context of modeling externally irradiated disks and relating such models to observable signatures and past observations. We also discuss the dynamical implications of an off-plane external heating source.

\subsection{Applications to observed systems and limitations}
\label{sub:observations}

Several disks in strongly irradiated environments have been observed to feature substructures of dynamical origin, including the ringed disks around SO~1274 and SO~844 \citep{huang-etal-2024} in $\sigma$~Orionis, the nonaxisymmetric disk around ISO-Oph 2 \citep{casassus-etal-2023}, and the seemingly warped proplyd 114-426 in Orion \citep{mccaughrean-etal-1998,miotello-etal-2012}. Our results could find applications---at least to an extent, given the underlying simplifications and assumptions laid out in Sect.~\ref{sec:physics}---in such systems, and encourage further exploration of this possibility with future work.

However, it is important to stress the limitations of both the mechanism explored here and the applicability of our models to the aforementioned systems. Possibly the most glaring difference between our model and the disks presented in \citet{huang-etal-2024} is the strength of the external irradiation flux, with our fiducial model being subject to a flux of $\sim\!10^5 \,G_0$, while the disks in $\sigma$~Orionis are exposed to a significantly weaker flux of $\sim\!10^{2.3}\,G_0$. While we expect that some level of nonaxisymmetric perturbations will be present even at lower fluxes, it is unclear whether the mechanism explored here can lead to prominent substructures within the lifetime of the disk, especially when considering mass loss due to external photoevaporation \citep[e.g.,][for $\sigma$~Orionis]{ansdell-etal-2017}.

To that end, we can qualitatively draw a parallel to our supplementary model with the external source placed at a distance of 1\,pc from the disk (left panel in Fig.~\ref{fig:appendix-heatmaps}), where we found that this reduction of the external flux to $\sim\!10^3\,G_0$ yielded weak perturbations in the gas surface density but no observable substructures in the dust over 100\,kyr. While this suggests that the mechanism explored here is not necessarily the dominant driver of the observed substructures in $\sigma$~Orionis, we note that a more detailed exploration of the parameter space in addition to more sophisticated modeling is needed to fully understand the relevance of this mechanism for the observed systems, as our models are not meant to be their direct analogs.
\vspace{-1em}

\subsection{3D effects and self-shielding}
\label{sub:3d-effects}

In this work we limited our models to a 2D, vertically integrated framework, assuming that the external heating source lies within the disk plane. While this approach enables a first look at the dynamical impact of external irradiation, it represents a rather ideal situation where the top-down symmetry of the disk is maintained. In practice, starlight from the external source could impinge on the disk at an arbitrary angle \citep[see, e.g.,][]{ricci-etal-2008}, shifting the location of the disk midplane as one hemisphere is preferentially heated. This could then induce significant vertical motion in the form of a warp in the disk, complicating its dynamical evolution \citep[see, e.g.,][]{rabago-etal-2024,kimmig-dullemond-2024,zhang-etal-2025} and driving likely observable features \citep[e.g.,][]{kimmig-villenave-2025}. In fact, it has been suggested that the seemingly warped structure in the Orion~114-426 disk \citep{miotello-etal-2012} could be induced by asymmetric external irradiation/photoevaporation. This possibility will be explored with followup work, using 3D radiation hydrodynamical simulations.

\citet{garate-etal-2024} further found that disk substructures can delay dust loss under external photoevaporation by enhancing dust trapping and shielding. Their perturbation strengths of $\Delta\Sigmag/\Sigma_0\sim\!0.25$ exceed ours ($\sim\!0.1$ at 100\,kyr, see Fig.~\ref{fig:radial-profiles}), but our structures are expected to grow with continued asymmetric heating \citep{ziampras-etal-2025b,zhu-etal-2025}. This makes it plausible that similar feedback could operate in our scenario.

\section{Summary}
\label{sec:summary}

In this work, we have explored the dynamical impact of external irradiation on protoplanetary disks with two-dimensional hydrodynamical simulations including dust--gas dynamics, radiation transport, and an asymmetric heating term due to an in-plane nearby massive star. Our results can be summarized as follows:
\begin{itemize}
	\item We have demonstrated for the first time that, if external irradiation can drive even modest asymmetric external heating of protoplanetary disks, this can lead to prominent dynamical effects within the disk, including the formation of pronounced spiral arms in gas, concentric rings and gaps in dust, and a noticeably lopsided redistribution of mass within the disk.
	\item The mechanism behind these dynamics is a direct analog of the shadow-induced spirals and rings studied in disks with a misaligned inner component shadowing the outer disk, with the primary difference being that here the disk itself is the source of the shadowing against the external heat source.
	\item These dynamics may lead to observable signatures in mm continuum emission, potentially related to those seen by ALMA in strongly irradiated regions such as $\sigma$ Orionis, although this connection remains tentative.
\end{itemize}

Our findings suggest that external irradiation, in addition to driving photoevaporative winds, if external irradiation can lead to asymmetric heating it can also have significant dynamical effects on the structure of protoplanetary disks in a manner that cannot be captured by axisymmetric models. This could have important implications for both disk evolution and planet formation in strongly irradiated environments, and encourages further investigation with future studies. In particular, given the complex radiative transfer and dynamics of the disk/wind system, which we have simplified here, it remains to be proven in future work whether this mechanism is expected to be significant for real systems.

\begin{acknowledgements}
	We thank the referee for their careful review and constructive comments that significantly improved the manuscript. AZ acknowledges funding by STFC grant ST/T000341/1 and ST/X000931/1, and from the European Union under the European Union's Horizon Europe Research and Innovation Programme 101124282 (EARLYBIRD). LQ and TJH acknowledge UKRI guaranteed funding for a Horizon Europe ERC consolidator grant (EP/Y024710/1). TJH also acknowledges a Royal Society Dorothy Hodgkin Fellowship. This research utilized Queen Mary's Apocrita HPC facility, supported by QMUL Research-IT (http://doi.org/10.52x81/zenodo.438045). The equipment was funded by BEIS capital funding via STFC capital grants ST/K000373/1 and ST/R002363/1 and STFC DiRAC Operations grant ST/R001014/1. Views and opinions expressed are those of the authors only. All plots in this paper were made with the Python library \texttt{matplotlib} \citep{hunter-2007}.
\end{acknowledgements}

\vspace{-2em}

\bibliographystyle{aa}
\bibliography{refs}

\newpage

\begin{appendix}

\section{Data availability}
Data from our numerical models are available upon reasonable request to the corresponding author.

\section{Physics and numerics}
\label{sec:physics-numerics}

Here we detail the physical and numerical framework of our simulations, highlighting the differences with respect to \citet{ziampras-etal-2025b}. We also describe our implementation of external irradiation heating, as well as our initial and boundary conditions.

\subsection{Physics}
\label{sec:physics}

We consider a razor-thin protoplanetary disk orbiting a star with mass $\Mstar=1\,\Msun$ and luminosity $\Lstar=1\,\Lsun$, such that the Keplerian frequency at distance $R$ is $\OmegaK=\sqrt{\G\Mstar/R^3}$, with $\G$ the gravitational constant. The disk mainly consists of gas with adiabatic index $\gamma=7/5$, mean molecular weight $\mu=2.353$, surface density $\Sigmag$, vertically integrated internal energy density $e$, and a velocity field $\velg$, with a pressureless dust component with surface density $\Sigmad$ and velocity $\veld$. Via the perfect gas closure relation we can define the vertically integrated pressure $P=(\gamma-1)e$, the isothermal sound speed $\cs=\sqrt{P/\Sigmag}$, and the temperature $T=\mu \cs^2/\Rgas$, with $\Rgas$ being the perfect gas constant. The disk is exposed to irradiation from an external source located at a distance $\dext$ from the central star, emitting a flux $\vecFext$ with bolometric luminosity $\Lext$.

Our physical framework largely follows that of \citet{ziampras-etal-2025b}, with a few modifications that we will highlight in this section. In particular, we solve the Navier--Stokes equations for both gas and mm-grain dust components, coupled via a drag term in the Epstein regime \citep{Weidenschilling-1977}. We further include heating via viscosity and irradiation (both internal and external), as well as cooling via surface losses and a simplified treatment of the in-plane radiative flux. With that in mind, the only differences between our current framework and that of \citet{ziampras-etal-2025b} are found in the energy equation, which here reads:
\begin{equation}
	\label{eq:navier-stokes-3}
	\D{e}{t}=-\gamma e\nabla\cdot\velg+Q_\mathrm{visc}+Q_\mathrm{irr} + Q_\mathrm{cool} + Q_\mathrm{rad}' + Q_\text{ext},
\end{equation}
where $Q_\text{visc}$, $Q_\text{irr}$, and $Q_\mathrm{cool}$ are the viscous heating, internal irradiation heating, and surface cooling terms, respectively (as defined in \citealt{ziampras-etal-2025b}). Here, $Q_\text{rad}'$ captures in-plane cooling via a simple $\beta$-cooling approach
\begin{equation}
	\label{eq:Q_rad}
	Q_\mathrm{rad}' = -\Sigmag\cv \frac{T - T_0}{\beta}\OmegaK, \quad T_0 = 90\,R_\text{au} ^{-3/7}\,\text{K},
\end{equation}
where $T_0$ is the initial temperature profile of the disk given by a balance between $Q_\text{irr}$ and $Q_\text{cool}$, and $\beta$ follows the prescription for in-plane cooling of \citet{ziampras-etal-2023a} with
\begin{equation}
	\label{eq:beta_cool}
	\beta = \frac{\OmegaK}{\eta}\left(H^2 + \frac{\lrad^2}{3}\right), \quad \eta = \frac{16\sigmaSB T^3}{3\kappaR\rhomid^2\cv}, \quad \lrad = \frac{1}{\kappaP \rho_\mathrm{mid}}.
\end{equation}
In the above, $\cv$ is the specific heat capacity at constant volume, $\sigmaSB$ is the Stefan--Boltzmann constant, $\kappaR$ and $\kappaP$ are the Rosseland and Planck mean opacities, respectively, with $\kappaR=\kappaP=\kappa$ given by the opacity model of \citet{bell-lin-1994}, $H=\cs/\OmegaK$ is the pressure scale height and $\rhomid=\Sigmag/(\sqrt{2\pi}H)$ is the midplane gas density. We note that this approach is mostly valid in the context of temperature perturbations \citep[e.g.,][]{miranda-rafikov-2020b,ziampras-etal-2023a}, which is the case in the model discussed here. Equation~\eqref{eq:Q_rad} then practically approximates radiative losses as such perturbations decay, and $T_0$ acts as an analytically available profile that corresponds to the equilibrium solution and effectively a temperature floor in the shadowed and illuminated disk sides, respectively.

Finally, $Q_\text{ext}$ captures heating from external irradiation, which we model as
\begin{equation}
	\label{eq:Q_ext}
	Q_\text{ext} = -\sqrt{2\pi} H\,\left(\nabla\cdot\vecFext\right)\big|_{z=0},\quad \vecFext = \frac{\Lext}{4\pi \dext^2} e^{-\tau}\hat{d},
\end{equation}
with $\tau$ the optical depth between the external source and a given point in the disk, and $\Lext=2\times10^5\,\Lsun$ by default. Assuming that $\bm{d}_\text{ext} = \dext \hat{x}$, that is, the external source is located in-plane along the x-axis, and that it lies far enough such that the rays can be considered parallel to each other when reaching the disk, we can write $\tau$ at a distance $x=R\cos(\phi)$ from the central star as
\begin{equation}
	\label{eq:tau_ext}
	\tau(x) = \int_{\dext}^{x} \kappa\,\rho\,\mathrm{d}x'.
\end{equation}
The opacity $\kappa$ used here is once again computed using the model of \citet{bell-lin-1994}. This choice of an infrared opacity is motivated by the fact that the external UV irradiation \citep[to which][would not apply]{bell-lin-1994} is attenuated in the wind before it penetrates the disk. Heating of the disk itself due to external irradiation hence comes from infrared emission from the warm photodissociation region (PDR) in the wind \citep{1980ApJS...44..403R, 2025MNRAS.537..598K,keyte-haworth-2026}, where for the sake of simplicity we assume that this corresponds to the full bolometric luminosity of $\Lext=2\times10^5\,\Lsun$ from the external source, here a nearby O-type star.

We stress that this approach is approximate---to the point of simplistic---compared to detailed radiative transfer, photochemistry and thermal balance, but captures the presence of a nonaxisymmetric heating rate (here, from an external source) while striking a reasonable balance between simplifying the complicated thermochemistry at the PDR and maintaining a parameterizable setup that nevertheless yields reasonable temperatures within the disk when compared to previous models \citep[e.g.,][]{haworth-2021}. As it is therefore unclear whether the resulting heating rate generally under- or overestimates the temperature output from a fully self-consistent thermochemical calculation, we believe that it is sufficient given the exploratory nature of this work. 
Nevertheless, this work should be viewed as asking what the dynamical impact is if an externally induced temperature asymmetry is established, and the important next step is to undertake full radiation hydrodynamics in a range of external irradiation scenarios (e.g., UV-to-bolometric flux ratios) to determine under what conditions such an asymmetry is genuinely produced. At present, given that our estimate of the heating rate comes from thermal reemission, which is intrinsically more isotropic than direct illumination, we expect that the level of nonaxisymmetry of the heating rate corresponds to an upper limit compared to a full, 3D radiative transfer model.

To compute $\tau$ in practice, we first constructed a 3D model of our disk on a Cartesian grid assuming vertical Gaussian stratification (i.e., $\rho(R,z) = \rhomid(R)\,e^{-z^2/2H^2}$ with a vertically isothermal profile), performed ray tracing from the external source to each cell to compute $\tau$, and then interpolated the resulting $\tau(x,y)$ heatmap back onto our 2D polar grid.
This procedure is carried out using a Python script once, at the beginning of each simulation, assuming that the vertical structure of the disk does not change significantly during the simulation runtime. As we will see in Sect.~\ref{sec:results}, this approximation becomes less accurate for very strong external irradiation due to both significant vertical expansion of the disk and the development of non-axisymmetric structures that would feed back onto the optical depth calculation. Nevertheless, as we intend for this work to be a first exploration of the dynamical impact of external irradiation on protoplanetary disks, we leave a more self-consistent treatment of the radiative transfer due to the external source for future studies.

\subsection{Numerics}
\label{sec:numerics}

We utilize the \pluto{} code \citep{mignone-etal-2007} to solve the above equations on a 2D polar grid $(R,\phi)$, spanning $R\in[5,500]$\,au and $\phi\in[0,2\pi]$, with $N_R\times N_\phi = 1048\times1427$ cells. Combined with a logarithmic radial spacing, this grid setup results in roughly square cells with a resolution of 16 cells per scale height at 100\,au. We use the HLLC \citep{toro-etal-1994} and \citet{leveque-2004} Riemann solvers for the gas and dust, respectively, with a second-order Runge--Kutta time integrator and piecewise linear spatial reconstruction, the FARGO algorithm \citep{masset-2000,mignone-etal-2012}, and the implicit drag module described in \citet{ziampras-etal-2025a}.

Our initial disk setup corresponds to a disk orbiting at Keplerian speed with a tapered surface density profile
\begin{equation}
	\label{eq:sigma_g}
	\Sigmag(R) = 1400\,\left(\frac{R}{R_0}\right)^{-1} e^{-R/R_0}, \quad R_0 = 100\,\text{au},
\end{equation}
for a total gas mass of $\approx0.1\,\Msun$, and a temperature profile given by Eq.~\eqref{eq:Q_rad}. The dust component corresponds to mm-sized grains with an initially uniform dust-to-gas ratio of 0.009, assuming that the dust-to-gas ratio in small, opacity-carrying grains is 0.001. We impose a density floor of $10^{-5}$\,g/cm$^2$ on both dust and gas to avoid numerical issues in low-density regions, although this has no impact on our results. We utilize wave-damping \citep{devalborro-etal-2006} and outflow boundary conditions at the inner and outer radial boundaries, respectively. Finally, we set $\Lext=2\times10^5\,\Lsun$ to mimic irradiation by a $\theta^1$~C Ori-like source located at $\dext=0.1$\,pc, and integrate for 100\,kyr, or 100 orbits at 100\,au.

\end{appendix}

\end{document}